\documentclass[a4paper]{jpconf}
\usepackage{graphicx}
\usepackage{bm}
\usepackage{cite}
\begin{document}
\title{Low-temperature specific heat for ferromagnetic and antiferromagnetic orders in CaRu$_{\bm{1-x}}$Mn$_{\bm x}$O$_{\bm 3}$}

\author{M Yokoyama$^{1,\dag}$, S Nakano$^1$, S Someya$^1$, T Nakada$^1$, N Wada$^1$,\\ H Kawanaka$^2$, H Bando$^2$, K Tenya$^3$, A Kondo$^4$ and K Kindo$^4$}
\address{$^1$Faculty of Science, Ibaraki University, Mito 310-8512, Japan}
\address{$^2$National Institute of Advanced Industrial Science and Technology, Tsukuba 305-8568, Japan}
\address{$^3$Faculty of Education, Shinshu University, Nagano 380-8544, Japan}
\address{$^4$Institute for Solid State Physics, The University of Tokyo, Kashiwa 270-8581, Japan}
\ead{$^\dag$makotti@mx.ibaraki.ac.jp}

\begin{abstract}
Low-temperature specific heat of CaRu$_{1-x}$Mn$_x$O$_3$ was measured to clarify the role of d electrons in  ferromagnetic and antiferromagnetic orders observed above $x=0.2$. Specific heat divided by temperature $C_p/T$ is found to roughly follow a $T^2$ function, and relatively large magnitudes of electronic specific heat coefficient $\gamma$ were obtained in wide $x$ range. In particular, $\gamma$ is unchanged from the value at $x=0$ (84 mJ/K$^2$ mol) in the paramagnetic state for $x \le 0.1$, but linearly reduced with increasing $x$ above $x= 0.2$. These features of $\gamma$ strongly suggest that itinerant d electrons are tightly coupled with the evolution of magnetic orders in small and intermediate Mn concentrations.   
\end{abstract}

\section{Introduction}
The relationship between anomalous charge transport and magnetism in the vicinity of metal-insulator transition is a longstanding subject in the physics of strongly correlated electron systems. The Ru-based metal CaRuO$_3$ is considered to have a paramagnetic ground state with itinerant characteristics of Ru 4d electrons \cite{rf:Cao1997,rf:Kiyama1998}. In CaRuO$_3$, magnetization is found to be significantly enhanced at low temperatures \cite{rf:Kiyama1999,rf:Felner2000}, while no long-range magnetic order is observed in microscopic measurements such as neutron scattering and NMR \cite{rf:Martinez1995,rf:Mukuda1999,rf:Yoshimura1999}. In addition, electrical resistivity $\rho$ shows an unusually large magnitude at high temperatures, corresponding to a very small mean-free pass of electron conduction comparable to lattice constants \cite{rf:Cao2008}. These features imply that CaRuO$_3$ involves instabilities for both magnetism and charge transport, which are enhanced by strong electronic correlation between itinerant Ru 4d electrons.

In the mixed compound CaRu$_{1-x}$Mn$_x$O$_3$ \cite{rf:Sugiyama1999,rf:Shames2004,rf:Markovich2006,rf:Taniguchi2008,rf:Mizusaki2009,rf:Kawanaka2009}, it is revealed that a ferromagnetic (FM) order evolves above $x=0.2$, and becomes stable in the intermediate $x$ range. Both spontaneous magnetization $M_0$ and Curie temperature $T_{\rm C}$ show maximum values at $x\sim 0.7$ ($M_0\sim 1\mu_{\rm B}/{\rm f.u.}$ and $T_{\rm C}=170\ {\rm K}$), followed by the reductions of them with further increasing $x$. Furthermore, an antiferromagnetic (AFM) spin arrangement with the G-type structure is observed in the FM ordered phase. Finally, at $x=1$ the FM order vanishes, and only the G-type AFM order appears below $T_{\rm N}=120\ {\rm K}$ \cite{rf:Wollan1955}. By doping Mn, the feature of $\rho$ also changes from metallic to less conductive above $x \sim 0.1$, but the magnitude of $\rho$ still remains in small values in the wide $x$ range, that is, smaller than $0.1\ \Omega$ cm even at low temperatures for $x \le 0.8$. At $x=1$, in contrast, $\rho$ dramatically increases with decreasing temperature, whose magnitude at low temperatures exceeds $10^6$ $\Omega\ {\rm cm}$ \cite{rf:Neumeier2000}. These properties indicate that doping Mn into CaRuO$_3$ modifies the itinerant electronic states due to strong correlations between Ru 4d and Mn 3d electrons, and it significantly affects the magnetic and transport properties. To clarify the variation of the d electronic states, we have performed the low-temperature specific-heat measurements for CaRu$_{1-x}$Mn$_x$O$_3$.

\section{Experiment Details}
Polycrystalline samples of CaRu$_{1-x}$Mn$_x$O$_3$ for $0 \le x \le 1$ were prepared by means of the conventional solid-state method. The mixture of appropriate amounts of CaCO$_3$, RuO$_2$ and MnO were first calcined at 800 $^\circ$C for 24 hours. They were shaped into pellets after careful mixing, and then sintered at 1200  $^\circ$C for 48 hours. By means of powder X-ray diffraction technique we checked that all the samples have a distorted perovskite structure (the GaFeO$_3$-type orthorhombic structure) \cite{rf:Sugiyama1999,rf:Shames2004,rf:Markovich2006,rf:Taniguchi2008,rf:Mizusaki2009,rf:Kawanaka2009} without any extrinsic phase. Specific heat $C_p$ was measured between 3 K and 275 K with a thermal-relaxation method using both a commercial system (PPMS: Quantum Design) and a hand-made equipment, where we used plate-shaped samples with the mass of about 4 mg and 40 mg, respectively. The thermal-relaxation curves for all the measurements were well definitive, and the $C_p$ data obtained from both equipments were consistent within the experimental accuracy. Ac-susceptibility was measured in the temperature range of 5-300 K to estimate the FM and AFM transition temperatures. Frequency and amplitude of the applied ac field were 180 Hz and $\sim 0.5\ {\rm Oe}$, respectively.

\section{Results and Discussion}
\begin{figure}[tbp]
\begin{center}
\includegraphics[keepaspectratio,width=0.75\textwidth]{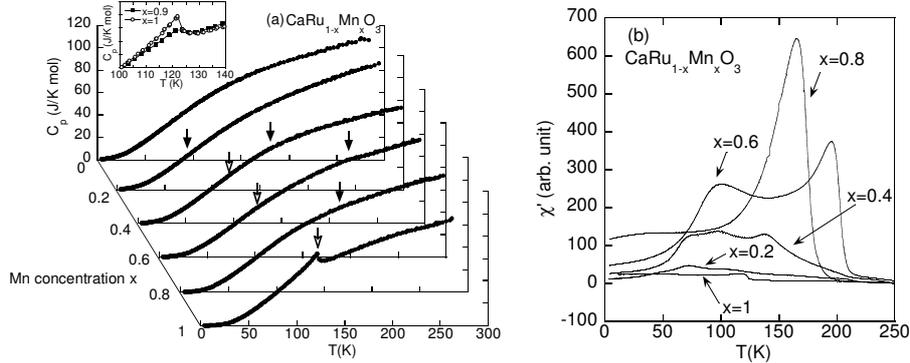}
\end{center}
\vspace{-15pt}
  \caption{
Temperature variations of (a) specific heat and (b) real part of ac susceptibility for CaRu$_{1-x}$Mn$_x$O$_3$. In (a), closed and open arrows indicate the onsets of FM and AFM orders, respectively, estimated from the peak position of the ac susceptibility. The types of magnetic order are determined from the previous reports on magnetization and neutron scattering measurements \cite{rf:Shames2004,rf:Markovich2006,rf:Taniguchi2008,rf:Mizusaki2009,rf:Kawanaka2009}. The inset of (a) shows an enlargement of $C_p$ around the AFM transition temperatures for $x=0.9$ and 1.
}
\end{figure}
Figure 1(a) shows temperature variations of the specific heat $C_p(T)$ for CaRu$_{1-x}$Mn$_x$O$_3$. Clear jumps ascribed to the AFM transition are observed at 122.7 K for $x=1$ and 125.1 K for $x=0.9$ (see inset of Fig. 1(a)), but no distinct anomaly is seen in the $C_p(T)$ curves for $x \le 0.8$. This contrasts with temperature dependence of the real part of ac susceptibility $\chi'(T)$ for $0.2 \le x \le 1$, where the FM and AFM transitions are indicated by the appearance of peaks (Fig.\ 1(b)). The transition temperatures estimated from $\chi'(T)$ are consistent with those determined by the dc magnetization and neutron scattering measurements \cite{rf:Shames2004,rf:Markovich2006,rf:Taniguchi2008,rf:Mizusaki2009,rf:Kawanaka2009}. The peaks seen in $\chi'(T)$ are somewhat broaden probably due to the distribution of the transition temperatures in the sample generated by doping. We consider this effect to be one of the reasons for the reduction of jumps in $C_p(T)$. On the other hand, it is interesting that similar reduction of anomaly is observed in the $C_p(T)$ data for Mn-doped SrRuO$_3$ \cite{rf:Yamaji2010}. SrRuO$_3$ is a ferromagnet with itinerant characteristics of Ru 4d electrons, and $C_p(T)$ exhibits a jump at $T_{\rm C}=162\ {\rm K}$. In the mixed compound SrRu$_{1-x}$Mn$_x$O$_3$, the anomaly at $T_{\rm C}$ in $C_p(T)$ is rapidly suppressed in the FM metallic phase for $x \le 0.3$, but a shoulder-like anomaly again appears in $C_p(T)$ at the AFM transition temperatures in the strongly insulating region for $x\ge 0.4$. As for CaRu$_{1-x}$Mn$_x$O$_3$, the electrical resistivity is much reduced for $x \le 0.8$. The disappearance of anomaly in $C_p(T)$ may thus be related to the small entropy change associated with the magnetic transitions in the small-resistivity region, as well as the broadening of the onset due to the inhomogeneity. 

\begin{figure}[tbp]
\begin{center}
\includegraphics[keepaspectratio,width=0.75\textwidth]{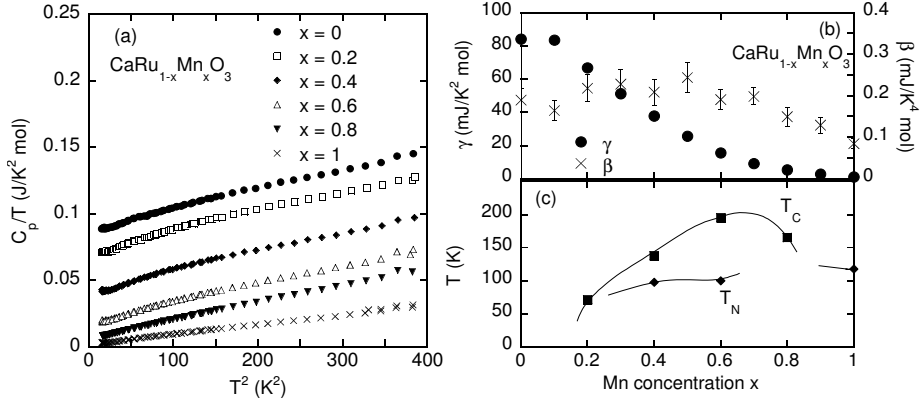}
\end{center}
\vspace{-10pt}
  \caption{
(a) Low-temperature specific heat divided by temperature plotted as a function of $T^2$, (b) electronic specific-heat coefficient $\gamma$ and lattice contribution on specific heat $\beta$, and (c) $T_{\rm C}$ and $T_{\rm N}$ estimated from the $\chi'(T)$ data for CaRu$_{1-x}$Mn$_x$O$_3$. The definition of parameters $\gamma$ and $\beta$ is described in the text. The solid lines in (c) are guides to the eye \cite{rf:Shames2004,rf:Markovich2006}.
}
\end{figure}
Displayed in Fig. 2(a) is low-temperature specific heat divided by temperature $C_p/T$ plotted as a function of $T^2$. $C_p/T$ for entire $x$ range are roughly proportional to $T^2$. The finite $C_p/T$ values for $T\to 0$ except at $x=1$ indicate that the Fermi-liquid excitation as well as phonon are mainly responsible for $C_p(T)$. However, $C_p/T$ in the small and intermediate $x$ range shows a very weak deviation from the $T^2$ function with a downward curvature below $T^2\sim 100\ {\rm K}^2$ ($T\sim 10\ {\rm K}$). It may be ascribed to the spin fluctuation generated in the magnetically ordered phase. The $C_p/T$ data are fitted with a function of $\gamma+\beta T^2$ to derive the electronic specific heat coefficient $\gamma$. In Fig. 2(b), we show $x$ variations of $\gamma$ and $\beta$ obtained from the best fit in the temperature range between 3 K and 10 K. For comparison, we also plot the $x-T$ phase diagram obtained from the $\chi'(T)$ data in Fig. 2(c). The $\gamma$ value for $x=0$ is estimated to be 84 mJ/K$^2$ mol, which is consistent with that reported previously \cite{rf:Kikugawa2009}. The magnitude of $\gamma$ does not change in the paramagnetic phase for $x \le 0.1$, but linearly decreases with increasing $x$ in the magnetically ordered region ($x\ge 0.2$). It is highly reduced and becomes $\sim 5\ {\rm mJ/K^2\ mol}$ or less for $x\ge 0.8$, where $T_{\rm C}$ decreases through the maximum at $x\sim 0.7$. On the other hand, $\beta$ is estimated to be 0.19 mJ/K$^4$ mol for $x=0$ that yields the Debye temperature of $\Theta_{\rm D}=370\ {\rm K}$. It is reduced in Mn-rich region and finally becomes 0.085 mJ/K$^4$ mol ($\Theta_{\rm D}=490\ {\rm K}$) at $x=1$. This feature seems to be consistent with the volume contraction in Mn-rich concentrations found by X-ray diffraction measurements \cite{rf:Shames2004,rf:Taniguchi2008}.
 
In CaRu$_{1-x}$Mn$_x$O$_3$, it is proposed that a double-exchange mechanism between Mn ions plays a crucial role in the evolution of the FM order in Mn-rich region, and in lower $x$ range an interplay of double-exchange and superexchange interactions between d electrons located at Ru and Mn ions may yield the formation of inhomogeneous FM and AFM orders \cite{rf:Shames2004,rf:Markovich2006}. On the other hand, the present $C_p$ investigation revealed that $\gamma$ shows relatively large values in small and intermediate $x$ ranges, while it is strongly reduced in Mn-rich region. In particular, $\gamma$ at $x=0.4$ is estimated to be 38 mJ/K$^2$ mol, which is almost the same magnitude as that for the ferromagnetic metal SrRuO$_3$ (33 mJ/K$^2$ mol). The large $\gamma$ values indicate that d electrons have itinerant characteristics in this $x$ range. Furthermore, we observed that $\gamma$ is nearly constant in the paramagnetic region, and it is then reduced when the FM order occurs above $x=0.2$. Such a variation of $\gamma$ cannot be simply attributed to an increase in the fraction of d electrons localized at Mn ions in the sample, but strongly suggests that the itinerant d electrons originating from both Ru and Mn are tightly coupled with the magnetic orders. This is supported by our recent magnetization and neutron-diffraction measurements for $x=0.4$, where the magnetization shows typical features on fluctuation expected from itinerant electron spins, and the neutron-diffraction experiments revealed the occurrence of the long-range magnetic orders \cite{rf:Kawanaka2009}.
 
\section{Summary}
We performed $C_p$ measurements for CaRu$_{1-x}$Mn$_x$O$_3$ ($0 \le x \le 1$). Low-temperature $\gamma$ value is estimated to be 84 mJ/K$^2$ mol for pure CaRuO$_3$, and it is unchanged by doping Mn in the paramagnetic region for $x \le 0.1$. In the magnetically ordered phase, $\gamma$ is linearly reduced with increasing $x$ for $0.2 \le x \le 0.6$, and then approaches $\sim 5$ mJ/K$^2$ mol or less for $x \ge 0.8$. The relatively large magnitude of $\gamma$ in the small and intermediate Mn concentrations suggests that d electrons have itinerant characteristics contrary to localized ones expected in Mn-rich concentrations, and they play important roles in the magnetic orderings.

\section*{Acknowledgment}
We thank Y. Nishihara and A. Yamaji for fruitful discussion. This work was carried out by the joint research in ISSP, the University of Tokyo, and partly supported by a Grant-in-Aid for Scientific Research on Innovative Areas ``Heavy Electrons" (No.23102703) from the Ministry of Education, Culture, Sports, Science and Technology of Japan.

\end{document}